\documentclass[12pt]{article}
\pdfoutput=1
\usepackage{pstricks,slashed}
\usepackage{color,verbatim}
\usepackage{amssymb,amsmath,bm,bbm,accents}
\usepackage{epsf,epsfig}
\usepackage{afterpage}
\usepackage{longtable}
\usepackage{cite}
\usepackage{latexsym,mathrsfs,dsfont}
\usepackage{graphics,graphicx}
\usepackage{url}
\usepackage{paralist}
\usepackage{bbold}

\usepackage[T1]{fontenc}
\usepackage[utf8]{inputenc}
\usepackage[english]{babel}
\usepackage{braket}
\usepackage{amsthm}
\usepackage{amsfonts}
\usepackage{float}

\newcommand{\be}{\begin{equation}}
\newcommand{\ee}{\end{equation}}
\newcommand{\bi}{\begin{itemize}}
\newcommand{\ei}{\end{itemize}}
\newcommand{\ba}{\begin{array}}
\newcommand{\ea}{\end{array}}
\newcommand{\bea}{\begin{eqnarray}}
\newcommand{\eea}{\end{eqnarray}}
\newcommand{\dd}{\displaystyle}
\newcommand{\nn}{\nonumber}

\newcommand{\ce}{\textsc{CE}}

\newcommand{\diff}{\mathrm{d}}
\newcommand{\azione}{\mathcal{S}}
\newcommand{\lagrangiana}{\mathcal{L}}

\newcommand{\numberset}{\mathbb}
\newcommand{\reale}{\numberset{R}}

\newcommand{\g}{\mathrm{g}}

\newcommand{\T}{\mathrm{T}}

\begin{document}

\begin{flushright} {BARI-TH/18-717}\end{flushright}

\medskip

\begin{center}
{\Large  Configurational Entropy  can disentangle  \\ \vspace*{0.3cm} conventional hadrons from exotica}
\\[1.0 cm]
{\large {P.~Colangelo$^a$  and F.~Loparco$^{a,b}$}
 \\[0.5 cm]}
{
$^a$Istituto Nazionale di Fisica Nucleare, Sezione di Bari, \\ Via Orabona 4, I-70126 Bari, Italy \\[0.1 cm]
$^b$Dipartimento Interateneo di Fisica ``M. Merlin", Universit\`a  e Politecnico di Bari, \\ via Orabona 4, 70126 Bari, Italy
}
\end{center}

\vskip 0.8cm

\begin{abstract}
\noindent
 We evaluate the Configurational Entropy  (CE) for  scalar mesons and  for $J^P=\frac{1}{2}^+$ baryons  in a holographic approach, varying the dimension of  boundary theory  operators and using the soft-wall dual model of QCD.  We find that hybrid and multiquark mesons  are characterized by an increasingly large \ce. A similar behavior  is observed for  $J^P=\frac{1}{2}^+$ baryons, where the {\ce}  of pentaquarks is larger than for three-quark baryons, for same radial number.  Configurational Entropy
seems relevant in disentangling conventional hadrons  from exotica.
\end{abstract}

\thispagestyle{empty}


\section{Introduction}
Color neutral multiquark/multigluon  hadrons  can exist in QCD, in principle  \cite{GellMann:1964nj}. Nevertheless,    by far the largest part of the observed hadrons can be  interpreted as  $\bar q q$ and $qqq$  "conventional" configurations. The observed meson spectrum can be constructed in terms of $\bar q q$ constituents with quantum numbers $J^{PC}$  attributed according to the simple rules of the quark model,  and the few exceptions,  candidates of $\bar q G q$  hybrids,  $\bar q \bar q  q q$ tetraquarks and $GG$ glueballs,  are the subject of disputes based on different interpretations \cite{exotic}. The same  situation  characterizes the baryon sector, where the observation of five-quark states has  been recently reported   only for systems comprising a heavy quark-antiquark pair \cite{Aaij:2015tga}. This must be contrasted with the plethora of  observed  conventional three-quark baryons and with the non-observation of pentaquarks made  of light quarks \cite{PDG}.

The prevalence of conventional ("ordinary") quark configurations over non-conventional ("exotic") multiquark/multigluon configurations has, of course, the origin  rooted in the nonperturbative dynamics of QCD.  Hints about its emergence come  from large $N_c$ arguments \cite{tHooft:1973alw}\footnote{Recent discussions can be found  in \cite{Lucha:2017gqq} an in references therein.}. Here we  explore the possibility that this regularity in the  spectrum can find an interpretation using the notion of  Configurational Entropy ({\ce}) for  hadrons.\footnote{Here we use the name {\it Configurational Entropy} usually adopted in the literature, although  {\it Configurational Complexity} has been recently proposed as the most appropriate name for the quantity we  analyze \cite{Gleiser:2018jpd}. }

The use of Configurational Entropy to characterize the energy density profile of spatially localized systems  and to measure the stored information 
 \cite{gleiser2012entropic,Gleisertot} 
 has been inspired by the information theory, where the Shannon Entropy can be attributed to a sequence of bits in a binary string \cite{Shannon:1948zz}. This notion has allowed to describe the properties of various systems, finding that those with lower CE are dominant   \cite{Casadio:2016aum}.
 In particular, in applications to the hadron spectroscopy, the properties of  high-spin light meson production  \cite{Bernardini:2016hvx},   the properties of the lowest spin glueballs    
 and of the meson Regge trajectories \cite{Bernardini:2016qit}, and  the features of  lowest-lying quarkonia at $T=0$ and  at finite temperature \cite{braga2018ads}
have been scrutinized  evaluating   {\ce}  in the framework of AdS/QCD models.
Here we wish to  compute {\ce}  in a holographic model, considering  fields dual to QCD  operators  interpolating multiquak/multigluon hadronic states, to scrutinize the regularities emerging when the number of  constituent quarks and gluons increases. 

A definition of Configurational Entropy  has been proposed in   \cite{gleiser2012entropic} 
for a system characterized by  a property described by a  square-integrable bounded function, the profile function ${\cal F}(x) \in L^2(\reale)$, having  Fourier transform 
\begin{equation}
F(k) = \int_{-\infty}^{+\infty} \diff x \, e^{-i k x} \, {\cal F}(x) \;
\end{equation}
 that satisfies the Plancherel relation
\begin{equation}
 \int_{-\infty}^{+\infty} \diff k \, |F(k)|^2 = \int_{-\infty}^{+\infty} \diff x \, |{\cal F}(x)|^2 \, .
\end{equation}
Starting from $F(k)$, the function
\begin{equation}
f(k) = \frac{|F(k)|^2}{\int_{-\infty}^{+\infty} \diff q \, |F(q)|^2} 
\end{equation}
represents the relative weights of the  various $k$ modes.
If $f(k)$  is non-periodic, the \emph{modal fraction} can be further defined,
\begin{equation}
\hat f (k) = \frac{f(k)}{f(k)_{\max}} \;,
\end{equation}
which is bounded, $0\leq \hat f (k) \leq 1$. $f(k)_{\max}$ is the maximum weight for the mode $k$.  The system Configurational Entropy  is defined as \cite{gleiser2012entropic} 
\begin{equation}
S_{\ce} = - \int_{-\infty}^{+\infty} \diff k \, \hat f (k) \, \ln \hat f(k) \;.
\end{equation}
This is a continuum generalization  of the Shannon information entropy,
\be\label{Shannon}
S_{SH}=-\sum_{\ell=1}^{n} p_\ell \ln p_\ell
\ee
defined for a system characterized by a discrete set of events $e_\ell$ ($\ell=1, \dots, n$), each one occurring with probability $p_\ell$ (with $\sum\limits_{\ell=1}^{n} p_\ell=1$) \cite{Shannon:1948zz}. 

To determine  {\ce} of meson and baryons,  a suitable square-integrable function ${\cal F}(x)$ for the various hadrons must be selected. Using  the soft-wall  holographic model of QCD such a function can be recognized, and  a computation for each hadron species can be carried out, as done in Sect.~\ref{scalar}  for  $J^{PC} = 0^{++}$  mesons and  in  Sect.~\ref{fermion}  for  $J ^P= \frac{1}{2}^+$  baryons.

\section{Scalar mesons}\label{scalar}

To compute {\ce}  for scalar mesons we use an approach based on a holographic model of QCD. AdS/QCD is a theoretical
 framework inspired by the AdS/CFT duality \cite{Maldacena:1997re} aimed, in the so-called bottom-up version\footnote{A review of AdS/CFT principles and applications can be found in \cite{Ammon:2015wua}.},  at constructing  higher dimensional  models sharing properties with QCD, namely in the hadron sector  \cite{Erlich:2005qh}.  We use  the soft-wall model   in   which linear  Regge trajectories for light mesons are recovered \cite{Karch:2006pv}. 
In Poincar\'e coordinates $x_M=(x_0,x_1,x_2,x_3,z)$, an AdS$_5$ manifold is described using the metric tensor
\begin{equation}
\label{tensoremetrico}
\g_{MN} = e^{2 A(z)} \, {\rm diag} (+1,-1,-1,-1,-1) = e^{2 A(z)} \, \eta_{MN} \; ,
\end{equation}
 with warp factor
\begin{equation}
\label{adsfactor}
A(z) = \ln{\left( \frac{L}{z} \right)} \; 
\end{equation}
and indices $M,N=0,1,2,3,5$.
$L$ is the AdS$_5$  radius.    To break the conformal symmetry and to construct 5D models for hadrons, in the  soft-wall model a quadratic background dilaton $\varphi$,  depending only on the bulk coordinate $z$,  is included in the  5D actions,
\begin{equation}
\label{dilatone}
\varphi(z) = c^2 \, z^2 \;
\end{equation}
whith $c$ a dimensionful parameter. 

Following gauge/gravity duality, a  scalar field $\phi(x,z)$  is associated to  each QCD gauge invariant  operator $O(x)$ with $J^{PC}=0^{++}$. It is described by  the 5D action
\begin{equation}
\label{azioneuguaglianza}
\azione_{5S}= \frac{1}{\kappa} \int \diff^5 x \,  \sqrt{|\g|} \, e^{-\varphi(z)} \, \lagrangiana(\phi, \partial \phi, \g) \; ,
\end{equation}
with
\be
\lagrangiana = \frac{1}{2} \, \g^{MN} \, \partial_M \phi \, \partial_N \phi - \frac{1}{2} \, m^2 \, \phi^2 \,\,\, 
\ee
and $\kappa$ a constant  making \eqref{azioneuguaglianza} adimensional.
The resulting equation of motion, in the general case of $d+1$-dimensions with
$\dd \sqrt{|\g|} = e^{(d+1)A(z)}$, reads:
\begin{equation}
\label{equaz1}
\phi'' + \left[ (d-1) A' - \varphi' \right] \phi' - \left[ m^2 \, e^{2A} + \eta^{\mu\nu} \, \partial_{\mu} \, \partial_{\nu} \right] \phi = 0 \; .
\end{equation}
The primes indicate derivatives with respect to the bulk coordinate $z$, and
the  indices $\mu, \nu$ run in $0,1,\dots, d-1$. 

According to the holographic dictionary  \cite{Ammon:2015wua},
$m^2$ is related to the dimension $\Delta$ of the boundary theory operator $O(x)$ dual to  $\phi(x,z)$,
\begin{equation}
\label{dizionarioscalare}
m^2 \, L^2 = \Delta (\Delta - d) \;.
\end{equation}
Hence, scalar operators of increasing $\Delta$ ($O=\bar q q$,   $\bar q \bar q q q$, etc.) are dual to fields $\phi$ of increasing mass $m$.

 Eq.~\eqref{equaz1} becomes an equation only in the variable $z$ using the ansatz
$\dd \phi(x,z) = e^{-i p \cdot x} \, a(p) \, Y(z)$:
\begin{equation}
\label{equaz2}
Y'' - \left[ \frac{d-1}{z} + 2 c^2 z \right] Y' - \left[ \frac{\Delta(\Delta-d)}{z^2} - \omega^2 \right] Y = 0 \;,
\end{equation}
with $p^2=\omega^2$. This can be expressed as a Schr{\"o}dinger-like equations after the (Bogolubov) transformation 
$Y(z) = B(z) \, X(z)$,
with 
$\displaystyle B(z) = e^{\frac{c^2 z^2}{2}} \, z^{\frac{d-1}{2}}$:
\begin{equation}
\label{schroedinger}
- X'' + V(z) \, X = \omega^2 \, X \; ,
\ee
and $ V(z)$ acting as a potential,
\be
 V(z) = (d-2) c^2 + \frac{d^2 - 1}{4 z^2} + c^4 z^2 + \frac{\Delta(\Delta-d)}{z^2} \;.
\end{equation}
The regular solution of Eq.~\eqref{schroedinger} \cite{Colangelo:2007pt}
\begin{equation}
X(z) = C \, e^{-\frac{c^2 z^2}{2}} \, z^{\frac{2 \Delta - d + 1}{2}} \, L \left( \frac{1}{4} \left( \frac{\omega^2}{c^2} - 2 \Delta \right), \Delta - \frac{d}{2}, c^2 z^2 \right) 
\end{equation}
is obtained in terms of the  Laguerre generalized  function $L$, with spectrum
\begin{equation}
\label{spettroscalari}
\omega_{n,\Delta}^2 = 4 c^2 (n+\frac{\Delta}{2}) \;
\end{equation}
for the  radial quantum number $n=0,1,\dots$\,.
For $d=4$ we have:
\begin{align}
X_{n,\Delta}(z) & = C_{n,\Delta} \, e^{-\frac{c^2 z^2}{2}} \, z^{\Delta - \frac{3}{2}} \, L_n^{\Delta-2} ( c^2 z^2 ) \,\,\, ,
\end{align}
and normalized eigenfunctions of \eqref{equaz2} 
\begin{equation}
\label{funzionescalare}
Y_{n,\Delta}(z) = \sqrt{\frac{2 \, \Gamma(n+1)}{\Gamma(n+\Delta-1)}} \, z^{\Delta} \, L_n^{\Delta-2} ( c^2 z^2 ) .
\end{equation}

Following the  proposal in \cite{Bernardini:2016hvx}, the Configurational Entropy for the various meson states can be computed using  the energy-momentum tensor obtained  from \eqref{azioneuguaglianza}, in correspondence to the regular 
  solution $\phi$ of the equation of motion. Applying the definition
\begin{equation}
\T_{MN} =  \frac{2}{\sqrt{|\g|}}  \frac{\partial \left( \sqrt{|\g|} \, e^{-\varphi} \, \lagrangiana(\phi,\partial \phi,\g) \right)}{\partial \g^{MN}} \;,
\end{equation}
which holds since the action does not depend on derivatives of the metric $g$, we have:
\begin{equation}
\T_{MN} =  e^{-\varphi} \, \left( \partial_M \phi \, \partial_N \phi - \frac{1}{2} \, \eta_{MN} \, \eta^{KQ} \, \partial_K \phi \, \partial_Q \phi + \frac{1}{2} \, \eta_{MN} \, m^2 \, e^{2A} \, \phi^2 \right)  \;.
\end{equation}
The $T_{00}$ component, computed with the on-shell solutions, reads:
\begin{equation}\label{mesT00}
\T_{00}^{n,\Delta}(z) = \Omega \, e^{- c^2 z^2} \, \left( \omega_{n,\Delta}^2 \, Y_{n,\Delta}(z)^2 - Y_{n,\Delta}'(z)^2 - \frac{\Delta(\Delta-4)}{z^2} \, Y_{n,\Delta}(z)^2 \right) \;,
\end{equation}
with $\Omega$ a factor irrelevant for the subsequent computation. This  is the profile function ${\cal F}(z)$ from which the  {\ce} is computed for scalar mesons, considering various radial numbers  $n$ and  different operator dimensions $\Delta$. 

\section{ $J^P=\frac{1}{2}^+$ baryons}\label{fermion}

There are different ways of describing baryons in a holographic framework \cite{Ammon:2015wua}. Here, to
 describe $J^P=\frac{1}{2}^+$ baryons we start from the 5D action for a fermion field $\Psi(x,z)$ in the AdS$_5$ background:
\begin{equation}
\label{azioneads}
\azione_{5F} = \frac{1}{k} \int \diff^4 x \, \diff z \, \sqrt{|g|} \,\left [ \frac{i}{2}  \overline{\Psi}  e^M_A \Gamma^A D_M \Psi  - \frac{i}{2} (D_M \Psi)^\dagger \Gamma^0  e^M_A \Gamma^A  \Psi- m \overline{\Psi} \Psi  \right ] .
\end{equation}
 $e_A^M$ are the 5D  inverse vielbein,  $\Gamma^M=e_A^M \Gamma^A$, and   $\Gamma^A = \left( \gamma^{\mu}, -i \gamma^5 \right)$ (with Greek index $\mu=0,1,2,3$, and $\gamma^5 = i \gamma^0 \gamma^1 \gamma^2 \gamma^3$) the $4 \times 4$ Gamma matrices in 5D.
The $\Gamma^M$ matrices obey the algebra
\begin{equation}
\left\lbrace \Gamma^M , \Gamma^N \right\rbrace = \left\lbrace e^M_A \, \Gamma^A , e^N_B \, \Gamma^B \right\rbrace = 2 \, \g^{MN} \;.
\end{equation}
The covariant derivative $D_M$ involves the spin connection
\begin{equation}
D_M  =  \partial_M - \frac{i}{4} \omega_{M A B} \, \Sigma^{AB} 
\end{equation}
with $\Sigma^{AB} = \frac{i}{2} \left[ \Gamma^A, \Gamma^B \right]$;  in our case we have
\be\label{derivata}
D_M  =  \partial_M + \frac{i}{2} A'(z) \, \Sigma_{M5}  \;.
\end{equation}

In the soft-wall model the background dilaton term $e^{-\varphi(z)}$  included in the action \eqref{azioneads}  can be removed  rescaling the spinor field
$\Psi(x,z)\to  e^{\frac{\varphi(z)}{2}} \Psi(x,z)$ \cite{Kirsch:2006he}. To obtain normalizable solutions of the EOM and a discrete spectrum, it has been proposed to modify  the mass term in Eq.~\eqref{azioneads}  \cite{Gutsche:2012wb},
\begin{equation}
\label{rotturasimmetria}
m \to m_z = m + J(z) \; ,
\end{equation}
with
\begin{equation}
\label{jfunction}
J(z) = \frac{c^2 \, z^2}{L} 
\end{equation}
and $c$ the same coefficient  appearing in the quadratic dilaton $\varphi$.

Separating the fermion field in left-handed (L) and right-handed (R)  components,
\begin{equation}
\label{chirale}
\Psi(x,z) = \Psi_L(x,z) + \Psi_R(x,z) \;,
\end{equation}
the solutions for the L and R  component can be written as
\begin{equation}
\label{ondapiana}
\Psi_{L/R}(x,z)  = e^{- i p \cdot x} \, u_{L/R}(p) \, F_{L/R}(z) \;,
\end{equation}
with $u_{L/R}(p)$ Dirac spinors in 4D momentum space and $F_{L/R}$  functions of  $z$. 
Using Eqs.\eqref{chirale} and \eqref{ondapiana},  with $p^{\mu} = (\lambda,\vec 0)$, two coupled equations for $F_{L/R}$ (for $d=4$) are obtained,
\begin{equation}\label{eqF}
\left( \partial_z - \frac{1}{2} \left( \frac{d}{z} + 2 c^2 z \right) \pm \frac{m_z \, L}{z} \right) u_{L/R}(p) \, F_{L/R} (z)= \pm \lambda \, u_{R/L}(p) \, F_{R/L}(z) \; .
\end{equation}
Such equations can be decoupled,
\begin{equation}
\label{disaccoppiate}
\left( \partial_z - \frac{1}{2} \left( \frac{d}{z} + 2 c^2 z \right) \mp \frac{m_z \, L}{z} \right) \left( \partial_z - \frac{1}{2} \left( \frac{d}{z} + 2 c^2 z \right) \pm \frac{m_z \, L}{z} \right) F_{L/R}(z) = - \lambda^2 \, F_{L/R}(z) \; ,
\end{equation}
obtaining
\bea
&&F_{L/R}''(z) - \left( \frac{d}{z} + 2 c^2 z \right) F_{L/R}'(z) \nn\\ &&+ \left( (d-1) c^2 + c^4 z^2 + \left( 1 + \frac{d}{2} \right) \frac{d}{2z^2} \mp \frac{m_z \, L}{z^2} - \frac{m_z^2 \, L^2}{z^2} \pm \frac{m'_z \, L}{z} + \lambda^2 \right) F_{L/R}(z) = 0 . \,\,\,\,\,\,\, 
\eea
The transformation
\begin{equation}
\label{bogoliubov}
F_{L/R}(z) = B(z) \, Y_{L/R}(z) \;,
\end{equation}
with  $B(z) = e^{\frac{c^2 z^2}{2}} \, z^{d/2}$,  brings to 
\begin{equation}
\label{erwin}
- Y''_{L/R} + U_{L/R} \, Y_{L/R} = \lambda^2 \, Y_{L/R} \;
\end{equation}
with 
\begin{equation}
U_{L/R}(z) = \pm \frac{m_z \, L}{z^2} + \frac{m_z^2 \, L^2}{z^2} \mp \frac{m_z' \, L}{z} \;
\end{equation}
acting  as a potential. Notice that the ansatz (\ref{rotturasimmetria},\ref{jfunction}) produces  a confining potential in the  $z \to \infty$ IR region.

The holographic dictionary  for fermion operators with spin $s = \frac{1}{2}$ connects the fermion mass $m$ and $\Delta$:
\begin{equation}
\label{dizioneriofermioni}
m L = \Delta - s  = \sigma \;.
\end{equation}
Solving \eqref{erwin} with  this condition  we have:
\begin{equation}
Y_{L/R}(z) = C_{L/R} \, e^{- \frac{c^2 z^2}{2}} \, z^{\frac{1}{2}\left( 1 + | 1 \pm 2 \sigma | \right)} \, L_n^{\frac{1}{2} |1 \pm 2 \sigma|}(c^2 z^2) \;,
\end{equation}
with   radial quantum number $n$ and
\begin{equation}
\lambda^2_{n,\sigma} = \begin{cases}
c^2 \left( 4 n + 1 + 2 \sigma + | 1 + 2 \sigma| \right) \qquad (L) \\
c^2 \left( 4 n + 3 + 2 \sigma + | 1 - 2 \sigma | \right) \qquad (R) .
\end{cases}
\end{equation}
For $\sigma \geq 1/2$ the spectrum  for L and R fields coincides \cite{Gutsche:2012wb}: 
\begin{equation}
\label{spettrofermioni}
\lambda^2_{n,\sigma} = 4 c^2 \left( n + \sigma + \frac{1}{2} \right) \;.
\end{equation}
The regular solutions
\begin{align}
Y_L^{n,\sigma}(z) & = C_L^{n,\sigma}  \, e^{- \frac{c^2 z^2}{2}} \, z^{\sigma+1} \, L_n^{\sigma + \frac{1}{2}}(c^2 z^2)  \nn \\
Y_R^{n,\sigma}(z) & = C_R^{n,\sigma} \, e^{- \frac{c^2 z^2}{2}} \, z^{\sigma} \, L_n^{\sigma - \frac{1}{2}}(c^2 z^2) \; ,
\end{align}
including the normalization factors,  correspond to
\begin{align}
\label{sinistra}
F_L^{n,\sigma}(z) & = \sqrt{\frac{2 \Gamma\left( n + 1 \right)}{\Gamma\left( n + \sigma + \frac{3}{2} \right)}} \, z^{\sigma+3} \, L_n^{\sigma + \frac{1}{2}}(c^2 z^2)  \\
\label{destra}
F_R^{n,\sigma}(z) & = \sqrt{\frac{2 \Gamma\left( n + 1 \right)}{\Gamma\left( n + \sigma + \frac{1}{2} \right)}} \, z^{\sigma+2} \, L_n^{\sigma - \frac{1}{2}}(c^2 z^2) \; .
\end{align}

We can now  compute the baryon  {\ce}   considering the energy-momentum tensor
\begin{equation}
\T_{MN} =  \frac{2}{\sqrt{|\g|}} \frac{\partial \left( \sqrt{|\g|} \, e^{-\varphi} \, \lagrangiana(\Psi, D \Psi, \g) \right)}{\partial \g^{MN}}
\end{equation}
with 
\begin{equation}
\lagrangiana =  \frac{i}{2}  \overline{\Psi}  e^M_A \Gamma^A D_M \Psi  - \frac{i}{2} (D_M \Psi)^\dagger \Gamma^0  e^M_A \Gamma^A  \Psi- m_z \overline{\Psi} \Psi  ,
\end{equation}
evaluated in correspondence to the on-shell solutions. Using Eq.~\eqref{ondapiana} we obtain:

\begin{equation}\label{barT00}
\T_{00}^{n, \sigma}(z) = \Lambda \, \frac{e^{-c^2 z^2}}{z} \, \lambda_{n,\sigma}^2 \, \left[ (F_L^{n,\sigma}(z))^2 + (F_R^{n,\sigma}(z))^2 \right] \;, 
\end{equation}
with the indices $n,\sigma$  in the spectrum \eqref{spettrofermioni} and in the solutions (\ref{sinistra},\ref{destra}), and $\Lambda$ a constant. This   is the  profile function ${\cal F}(z) $ that we use to compute the baryon {\ce}.

\section{Configurational Entropy for $J^{PC}=0^{++}$ and $J^P=\frac{1}{2}^+$ hadrons}
\label{results}

We can now  evaluate the Configurational Entropy of  $J^{PC}=0^{++}$ mesons and  $\dd J^P=\frac{1}{2}^+$ baryons starting from Eqs.~\eqref{mesT00} and \eqref{barT00}.
For scalar mesons, QCD gauge invariant operators with different (canonical) dimension\footnote{Since radiative effects are not taken into account in this description,   operator anomalous dimensions are not considered.}  $\Delta$ and  $J^{PC}=0^{++}$ can be constructed in terms of quark and gluon fields:
\be
 O=\bar q q , \hskip 0.5cm  O= Tr\left[\frac{\alpha_s}{\pi}  G_{\mu \nu}G^{ \mu \nu} \right] ,  \hskip 0.5cm  O= \bar q g_s \sigma_{\mu \nu}G^{\mu \nu}q ,    \hskip 0.5cm  O= \bar q \bar q q q,
\label{scalop}
\ee 
and so on. Analyses based on, e.g., QCD sum rules show that such operators  have large vacuum-particle matrix elements with meson states  interpreted in terms of  constituent quark and gluons, hence  conventional  mesons,  scalar glueball,   hybrid scalar mesons and scalar tetraquarks, respectively. Therefore, they can be used to investigate the regularities in {\ce}. The fields dual to the QCD operators are characterized by a different mass $m$,  and have already  been studied  in the holographic framework \cite{Colangelo:2007pt}.

\begin{figure}[t!]
\centering
\includegraphics[scale=0.72]{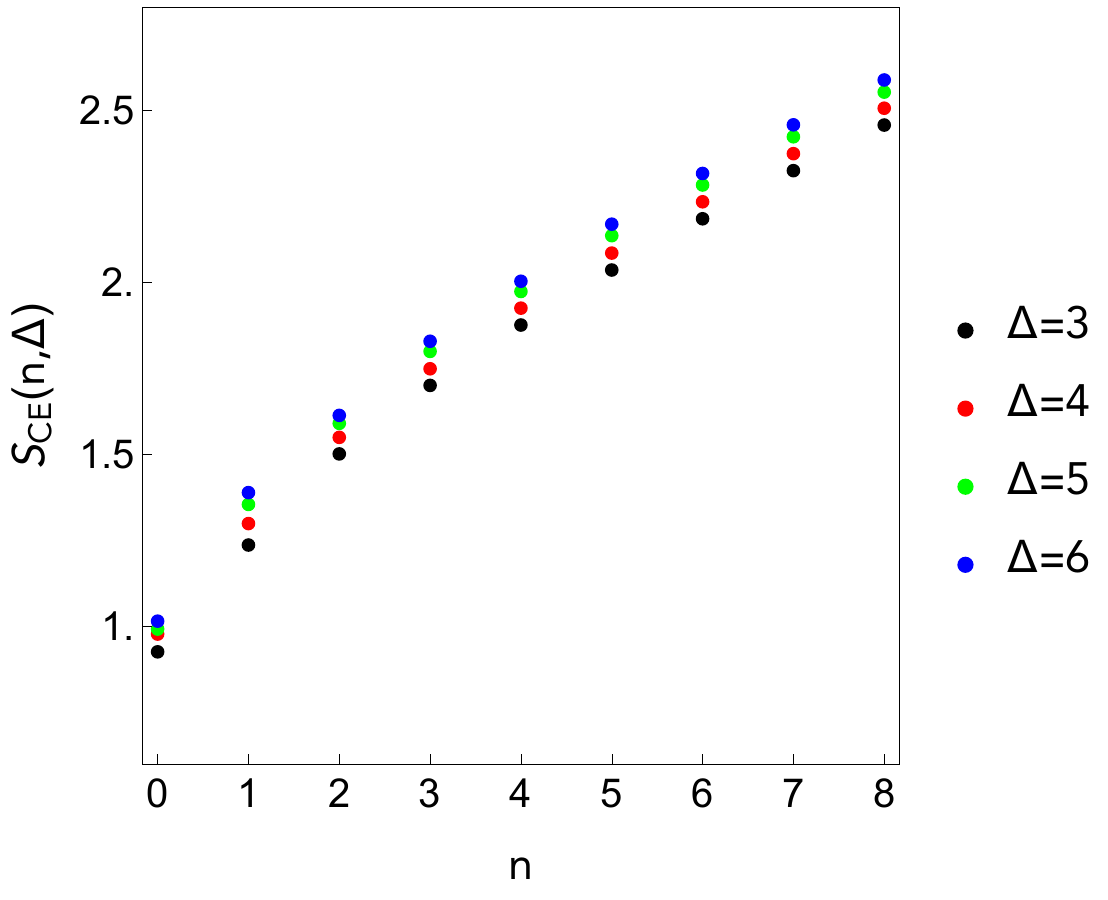} 
\vspace*{-0.2cm}
\caption{\baselineskip 10pt Configurational Entropy of  $J^{PC} = 0^{++}$ mesons  described by the QCD operators in Eq.~\eqref{scalop} with different $\Delta$.  $n$ is the radial quantum number. }
\label{scalarCE}
\end{figure}

Using the expression \eqref{mesT00},  the {\ce} can be computed for each value of $\Delta$ and  radial number $n$. The  result, obtained setting $c=1$,  is found to increase  with $n$, as shown in   Fig.~\ref{scalarCE}:  meson states corresponding to higher radial numbers  have  larger complexity content, as also observed in \cite{Bernardini:2016hvx,Bernardini:2016qit,braga2018ads}. 

\begin{figure}[t!]
\centering
\includegraphics[scale=0.6]{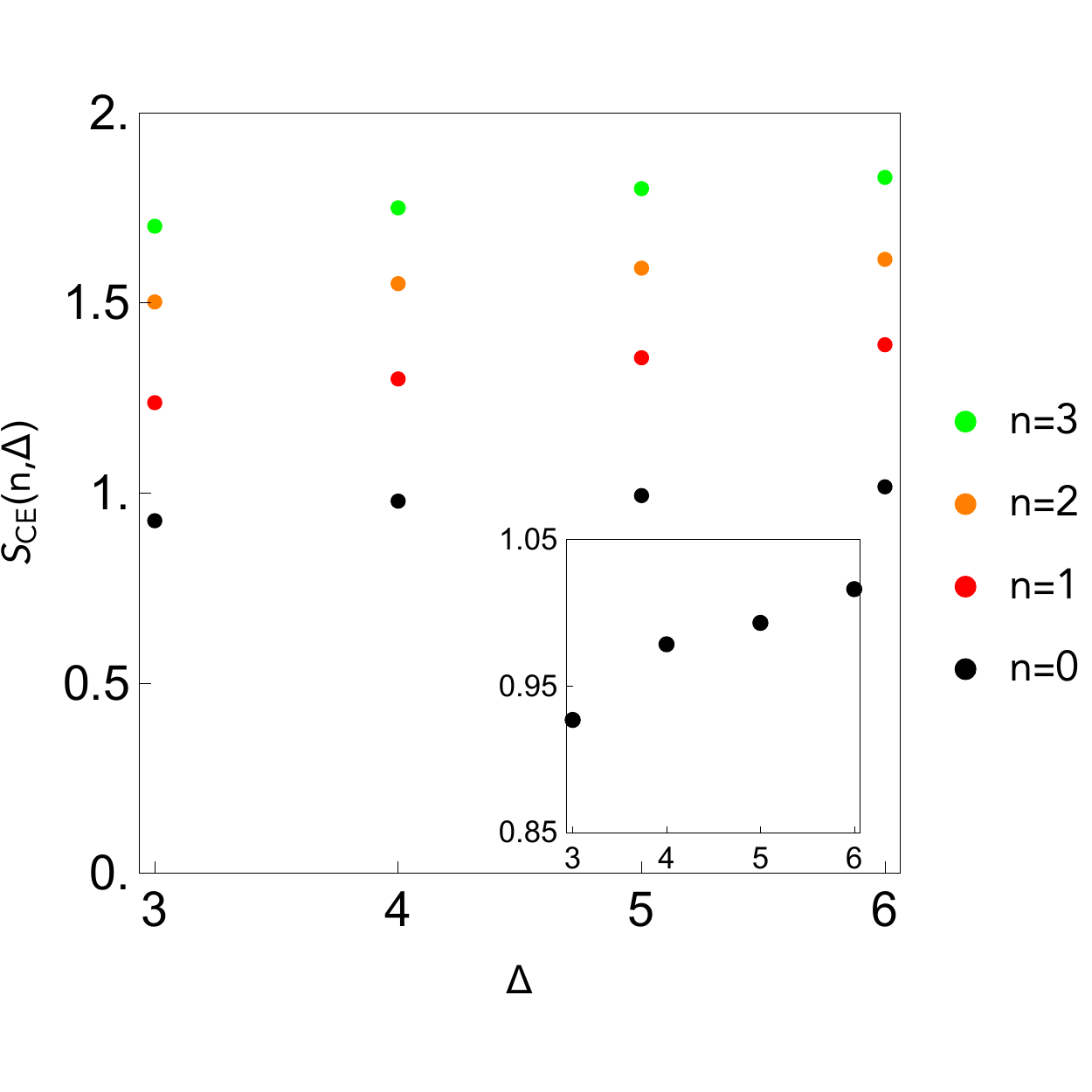}
\vspace*{-0.2cm}
\caption{\baselineskip 10pt Configurational Entropy of  scalar mesons  as in Fig.~\ref{scalarCE},  plotted versus the QCD operator dimension $\Delta$.  The results for the radial number $n=0$ are magnified in the inset.}
\label{scalarCE1}
\end{figure}

We are concerned with the dependence on the mass of the dual field, therefore on the operator dimension $\Delta$.  {\ce}
increases with the  dimension, indicating that states with a larger number of constituents
are characterized by a larger Configurational Entropy,  a behaviour  shown in Fig.~\ref{scalarCE1}. For the same value of $n$,  the scalar glueball  has larger {\ce}  than $\bar q q$ mesons.

Similar results are obtained for  $\dd J^P=\frac{1}{2}^+$ baryons. In this case, the different operator dimensions $\Delta$ correspond to various $\sigma$ values in Eq.~\eqref{dizioneriofermioni}. 
Examples of operators interpolating ordinary baryons and pentaquarks are the Ioffe's currents \cite{Ioffe:1981kw}
\bea
J^N(x)&=&\epsilon_{abc} \left(u^{aT}(x) {\cal C}\gamma_\mu u^b(x)\right)\gamma_5 \gamma^\mu d^c(x) \nn \\
J^{\prime N}(x)&=&\epsilon_{abc} \left(u^{aT}(x) {\cal C}\sigma_{\mu\nu} u^b(x)\right)\gamma_5 \sigma^{\mu \nu} d^c(x) \label{baryon-curr}
\eea
for the proton (with $\cal C$  the charge conjugation matrix and $a,b,c$ color indices), and
\bea
J^P(x)&=&\epsilon^{abc}\epsilon^{def}\epsilon^{cfg}\left(u^T_a(x) {\cal C}d_b(x)\right) \left(u^T_d(x) {\cal C}\gamma_5 d_e(x)\right) {\cal C} \bar s^T_g(x) \nn \\
J^{\prime P}(x)&=&-\epsilon^{abc}\epsilon^{def}\epsilon^{cfg} s^T_g(x) {\cal C} \left(\bar d_e(x)\gamma_5  {\cal C} \bar u^T_d(x)\right) \left( \bar d_b(x) {\cal C} \bar u^T_a(x)\right) 
\eea
for a light pentaquark with strangeness \cite{Sugiyama:2003zk}.
 
 \begin{figure}[b!]
\centering
\includegraphics[scale=0.72]{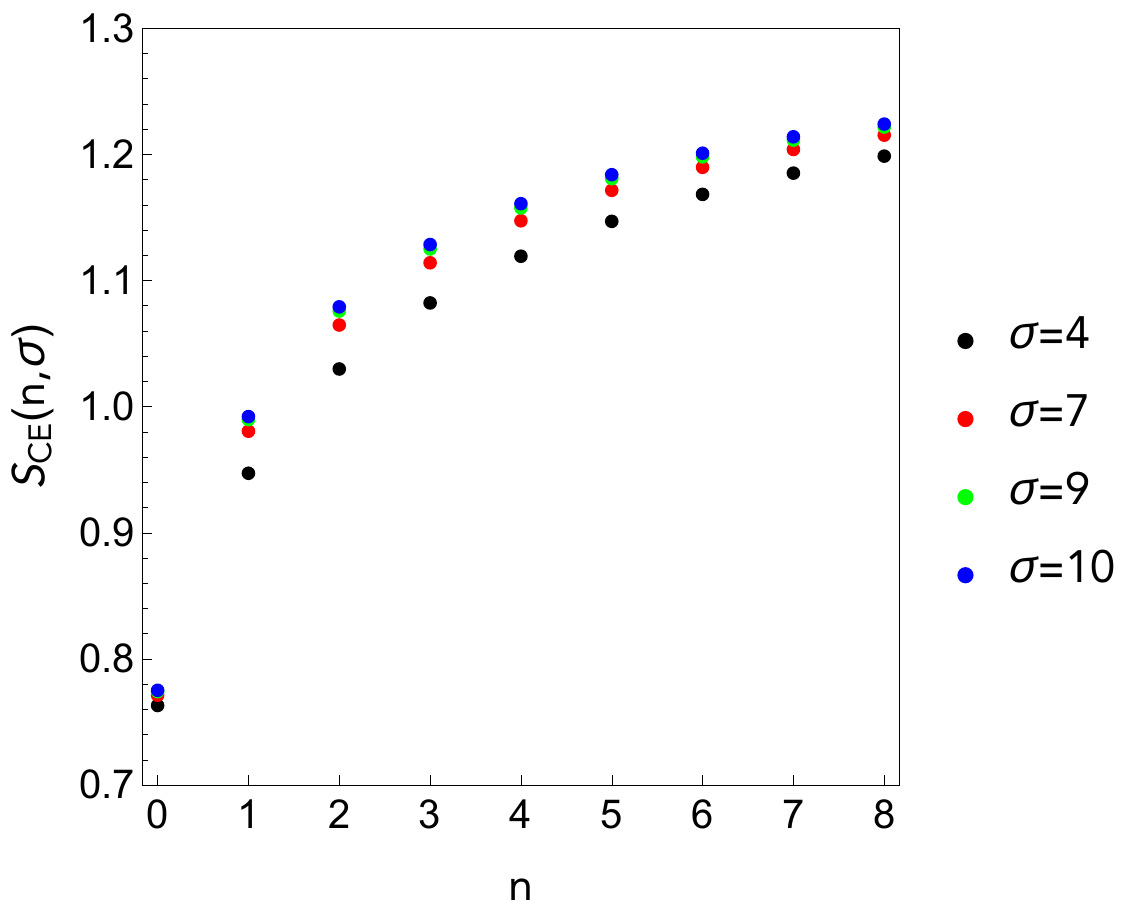} 
\vspace*{-0.2cm}
\caption{\baselineskip 10pt Configurational Entropy of  $J^P = \frac{1}{2}^+$ baryons described by QCD operators with different $\sigma$.  $n$ is the radial  number. }
\label{fermionCE}
\end{figure}

\begin{figure}
\centering
\includegraphics[scale=0.6]{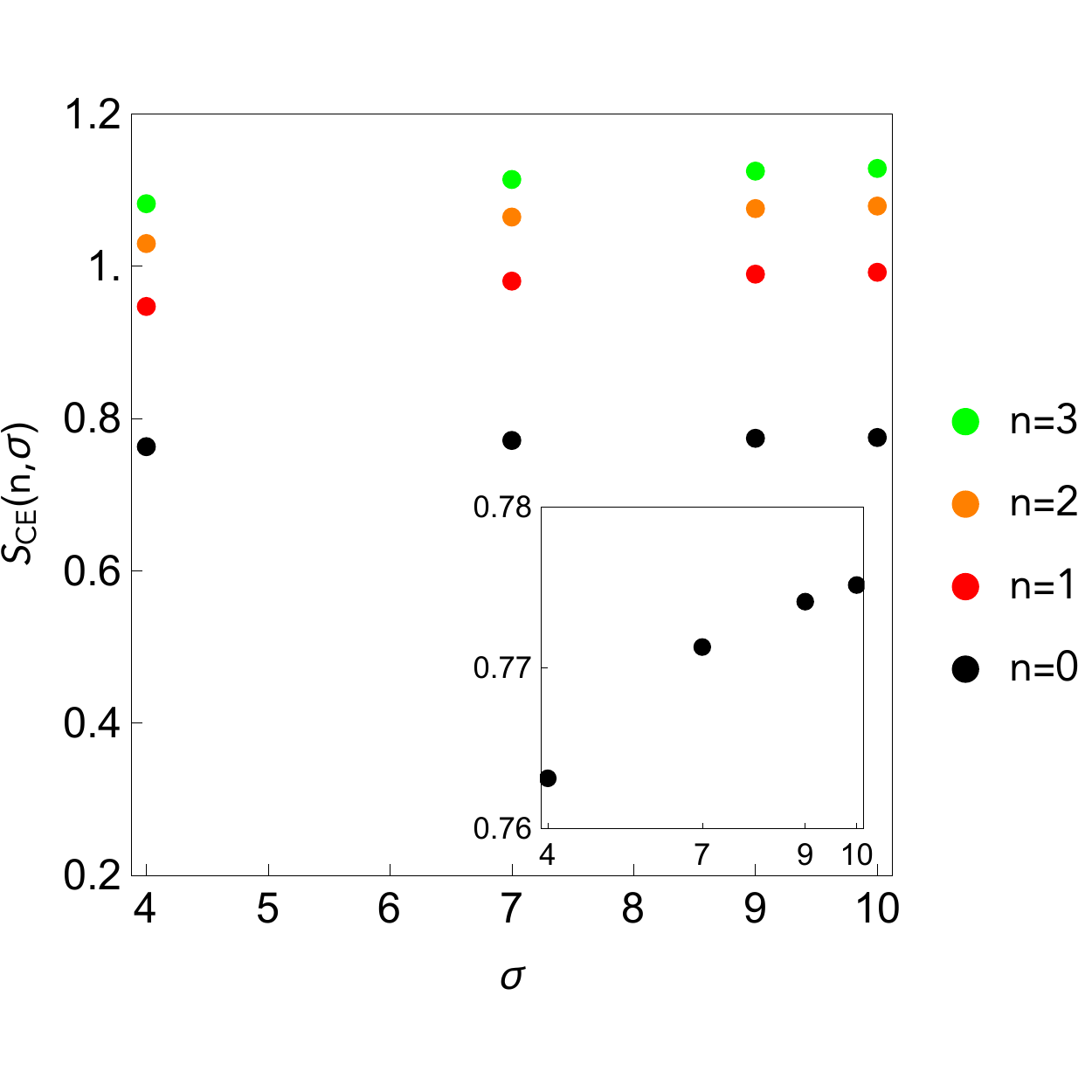}
\vspace*{-0.2cm}
\caption{\baselineskip 10pt Configurational Entropy of  $J^P = \frac{1}{2}^+$ baryons as in Fig.~\ref{fermionCE}, for QCD operators with different $\sigma$.    The results for radial number  $n=0$ are magnified in the inset.}
\label{fermionCE1}
\end{figure}

The  operators involve an increasing number of quark and gluon fields, and  their vacuum-particle matrix elements are found to be large in case of three-quark conventional hadrons ($\sigma=4$) and pentaquarks ($\sigma=7$).  The calculation of
{\ce} starting form Eq.~\eqref{barT00} gives again an increasing behaviour versus the radial number $n$, Fig.~\ref{fermionCE}. Moreover, increasing $\sigma$, the Configurational Entropy is also found to increase,  as shown in Fig.~\ref{fermionCE1}, and takes the minimum value for conventional baryons. 

\section{Conclusions}

The Configurational Entropy accounts for  information contained in the profile function.  Using the soft-wall  holographic model of QCD, it can be evaluated choosing, as a profile function, the function $T_{00}(z)$  computed in correspondence to the regular solutions of the equations of motion  for fields dual to QCD operators involving several quark and gluon fields. In addition to the increase of {\ce} with the radial number $n$, we found an increase, both in the scalar meson and in the baryon case,   with the operator dimension  related to the number and  kind of quark and gluon fields the QCD operator is made of. A possible interpretation of  this regularity is  that  smaller complexity characterizes  ordinary meson and baryon configurations.
 Configurational Entropy can be relevant in disentangling hadronic states when  the number and kind of their constituents is varied,  hence conventional hadrons from exotica.

\vspace*{1.cm}
\noindent {\bf Acknowledgements.}
We thank F. De Fazio, F. Giannuzzi and S. Nicotri for discussions. This study has been  carried out within the INFN project (Iniziativa Specifica) QFT-HEP.


\end{document}